# Strategies for protecting of multilayer networks from group and system-wide targeted attacks

Olexandr Polishchuk

Laboratory of Modeling and Optimization of Complex Systems
Pidstryhach Institute for Applied Problems of Mechanics and Mathematics, National Academy of Sciences of Ukraine,
Lviv, Ukraine
od_polishchuk@ukr.net

**Abstract** – *On the basis of structural model of intersystem interactions, the main local and global structural characteristics of nodes of the multilayer network (MLN) are determined. The notions of weighted and binary aggregate-networks of MLN are introduced and integral indicators of the importance of nodes of the multilayer network are determined. The effectiveness of application of aggregate-networks for solving a number of practically important problems of the theory of complex networks is shown. To determine the most important components of intersystem interactions structure, the concepts of p-core of multilayer network and kag-core of its weighted aggregate-network are introduced. The effective scenarios of successive and simultaneous group attacks on the structure of multilayer network systems have been built based on the use of these cores. Indicators of the importance of separate layers of MLN are introduced and approaches to building scenarios of successive and simultaneous system-wide attacks on the structure of intersystem interactions are formulated.*

**ВСТУП**

У статті [1] були проаналізовані основні види негативних внутрішніх та зовнішніх впливів на складні мережеві системи (МС) та процеси міжсистемних взаємодій. Серед таких впливів насамперед були виділені цілеспрямовані атаки та нецільові ураження складних систем, які можуть мати локальний, груповий або загальносистемний характер та бути спрямованими на ураження як структури, так і процесу функціонування мережевих та багатошарових мережевих систем (БШМС) різних типів. Натепер однією з основних стратегій захисту, яка розробляється в межах теорії складних мереж (ТСМ), є побудова так званих сценаріїв цілеспрямованих атак на структуру МС, які полягають у виділенні за певними ознаками (центральностями різних типів) найважливіших вузлів мережі та послідовному їх вилученні зі складу цієї структури [2, 3]. Основним недоліком таких сценаріїв є неоднозначність вибору типу центральності, яких існують десятки, та послідовне ураження вузлів МС під час якого система має змогу перерозподілити функції уражених елементів між тими, що залишились у складі структури [4]. Очевидно, що одночасні групові та загальносистемні атаки, під час яких відразу вражаються найважливіші складові структури або МС загалом, є набагато небезпечнішими та складнішими з погляду захисту та подолання наслідків цих атак. Прикладами таких цілеспрямованих уражень стали масовані ракетні удари по нафтобазах України, які призвели до суттєвих труднощів із постачанням пального у червні 2022 року, атаки на її енергосистему протягом жовтня–листопада 2022 року, унаслідок яких без електроенергії, водопостачання, зв'язку та Інтернету залишились найбільші міста та більшість регіонів країни, а також блокування роботи морських та аеропортів. Слід також враховувати, що локальні негативні впливи можуть переростати у групові і далі у загальносистемні, як це сталося з розгортанням Covid-19 у світову пандемію [5]. Що стосується дослідження уражень структури міжсистемних взаємодій, то натепер у ТСМ вони загалом обмежуються взаємозалежними (interdependent) багатошаровими мережами (БШМ), тобто ієрархічно-мережевими структурами із лінійною моделлю управління [6, 7], адже більшість складних систем, які функціонують у людському соціумі (промислових, фінансових, державних, військових тощо) мають саме таку структуру [8]. Однак, і у фізичному світі, і в людському суспільстві існує чимало інших типів структур міжсистемних взаємодій, зокрема частково покриті БШМ [9], ураження яких може призвести до катастрофічних наслідків.

Будь-яка реальна велика складна система не може захистити усі свої елементи та процес взаємодії з іншими системами [10, 11]. Натепер критерії вибору складових, які поєднують найважливіші за тими чи іншими ознаками вузли та ребра МС, особливо у випадку вироблення стратегій захисту від цілеспрямованих атак на переважну більшість видів БШМ, які описують структуру міжсистемних взаємодій, практично не вироблені [12]. Водночас виділення таких складових сприяє не лише кращому розумінню процесів, які перебігають у багатошарових мережевих системах, але й подоланню проблеми складності системних досліджень [13, 14].

**Мета статті** – розроблення на підставі структурних моделей монопотокових частково покритих багатошарових мережевих систем ефективних сценаріїв групових та загальносистемних цілеспрямових атак на структуру міжсистемних взаємодій у них.

**СТРУКТУРНА МОДЕЛЬ МІЖСИСТЕМНИХ ВЗАЄМОДІЙ**

Структурна модель міжсистемних взаємодій описується багатошаровими мережами [1, 15] та відображається у вигляді

$$G^M = \left( \bigcup_{m=1}^{M} G_m, \bigcup_{\substack{m,k=1 \\ m \neq k}}^{M} E_{mk} \right), \qquad (1)$$

де $G_m = (V_m, E_m)$ визначає структуру $m$-го мережевого шару БШМ; $V_m$ – множина вузлів мережі $G_m$; $E_m$ – множина зв'язків мережі $G_m$, $E_{mk}$ – множина зв'язків між вузлами множин $V_m$ та $V_k$, $m \neq k$, $m, k = \overline{1, M}$, де $M$ – кількість шарів БШМ. Множину

$$V^M = \bigcup_{m=1}^{M} V_m$$

називатимемо загальною сукупністю вузлів БШМ, $N^M$ – кількість елементів $V^M$.

Багатошарова мережа $G^M$ повністю описується матрицею суміжності $\mathbf{A}^M = \{\mathbf{A}^{km}\}_{m,k=1}^{M}$, у якій значення $a_{ij}^{km} = 1$, якщо існує ребро, яке з'єднує вузли $n_i^k$ та $n_j^m$, та $a_{ij}^{km} = 0$, $i, j = \overline{1, N^M}$, якщо такого ребра немає. При цьому блоки $\mathbf{A}^{mm}$ описують структуру внутрішньо шарових взаємодій у $m$-му шарі, а блоки $\mathbf{A}^{km}$ – структуру взаємодій між $m$-тим та $k$-тим шарами БШМ, $m \neq k$, $m, k = \overline{1, M}$. Для спрощення викладу вважатимемо, що ребра структури $G^M$ є неорієнтованими. Блоки $\mathbf{A}^{km} = \{a_{ij}^{km}\}_{i,j=1}^{N^M}$, $m, k = \overline{1, M}$, матриці $\mathbf{A}^M$ визначаються для загальної сукупності вузлів БШМ, тобто знімається проблема координації номерів вузлів у випадку їх незалежної нумерації для кожного шару. У цій статті ми розглядаємо випадок так званих частково покритих (partially overlapped, ЧП) багатошарових мереж у яких перетин множин вузлів $V_m$, $m = \overline{1, M}$, є непорожнім та міжшарові зв'язки можливі лише між вузлами з однаковими номерами із загальної сукупності вузлів БШМ (рис. 1) [16, 17]. Подібні структури міжсистемних взаємодій мають загальна транспортна система світу або окремої країни, транспортні системи великих міст, лінгвістична мережа та багато інших [18-20].

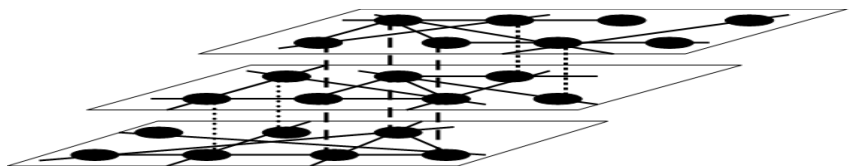

Рис. 1. Приклад структури частково покритої багатошарової мережі

З функціонального погляду це означає, що кожний вузол БШМ є елементом багатьох систем, у кожній з яких він реалізує одну функцію, але різними способами (населений пункт, як вузол залізничної, автомобільної, авіаційної та водної транспортних систем). Водночас існують і інші типи монопотокових міжсистемних взаємодій, у яких кожний вузол певного шару може мати множинні зв'язки з різними вузлами інших шарів БШМ (системи управління, поштові системи та системи телефонного зв'язку тощо) [21, 22].

# ЛОКАЛЬНІ ТА ГЛОБАЛЬНІ ХАРАКТЕРИСТИКИ ЕЛЕМЕНТІВ БАГАТОШАРОВИХ МЕРЕЖ

Під локальними характеристиками складової системи зазвичай розуміють кількісні показники, які описують той або інший аспект її взаємодії з безпосередньо пов'язаними (суміжними) складовими [23]. Глобальними вважаються характеристики складової системи, які описують кількісні показники її взаємодії з усіма іншими складовими цієї системи, або властивості системи загалом [24] . Усі локальні (вхідний та вихідний ступені, коефіцієнти кластеризації та розгалуження, розподіл кінців зв'язків і т. ін.) та глобальні (центральності різних типів, доступність та очікувана сила тощо) характеристики вузла у $m$-му шарі [25] вважатимемо його локальними внутрішньошаровими характеристиками у БШМ. Визначимо найважливіші для аналізу міжсистемних взаємодій локальні характеристики елементів частково покритої багатошарової мережі. Зазвичай локальною або глобальною характеристикою вузла БШМ у теорії складних мереж (ТСМ) вважається вектор його локальних або глобальних характеристик в окремих шарах системи [26, 27]. Наприклад, ступінь $\mathbf{d}_i$ вузла $n_i$ у БШМ визначається, як

$$\mathbf{d}_i = \{d_i^m\}_{m=1}^{M}, \quad d_i^m = \sum_{j=1}^{N^M} a_{ij}^{mm},$$

у якій значення $d_i^m$ визначають ступінь вузла $n_i$ загальної сукупності вузлів у $m$-му шарі, $i = \overline{1, N^M}$, $m = \overline{1, M}$. Такий підхід є цілком виправданим під час дослідження багатовимірних БШМС [28, 29], оскільки кількість зв'язків вузла у певному шарі таких БШМ загалом не залежить від кількості його зв'язків у інших її шарах. Однак, під час дослідження властивостей монопотокових ЧП БШМС локальні та глобальні структурні характеристики можна визначати не лише для вузлів окремих шарів, але й для сукупності міжшарових взаємодій загалом.

Оскільки зв'язки-петлі у шарах БШМ виключаються, тобто діагональні елементи матриць $\mathbf{A}^{mm}$ є нульовими, то параметр

$$s_{ij} = \sum_{m=1}^{M} a_{ji}^{mm},$$

за аналогією з відповідними поняттями теорії зважених мереж [30, 31] визначає силу взаємозв'язку або кількість способів реалізації цього зв'язку між вузлами $n_i$ та $n_j$ у багатошаровій мережі. Тоді параметр

$$s_i = \sum_{j=1}^{N^M} s_{ji} \tag{2}$$

визначає сумарну або агрегат-силу взаємозв'язків вузла $n_i$ у БШМ та є узагальненням поняття центральності за ступенем, яке визначається для вузлів бінарних мереж. Параметр

$$d_i = \sum_{j=1}^{N^M} \frac{s_{ij}}{\max(1, s_{ij})} \qquad (3)$$

визначає сумарну кількість зв'язків вузла $n_i$ у всіх шарах БШМ або агрегат-ступінь вузла $n_i$ у багатошаровій мережі відповідно. Іншими словами, значення $d_i$ визначає кількість суміжних із $n_i$ вузлів у БШМ загалом. Якщо агрегат-ступінь вузла $n_i$ визначає лише кількість суміжних вузлів цього вузла у всіх шарах БШМ, то агрегат-сила взаємозв'язку враховує число таких зв'язків у різних шарах, тобто варіантів взаємодій між вузлом $n_i$ та суміжними із ним вузлами у багатошаровій мережі.

Оскільки ми розглядаємо монопотокові системи у яких міжшарові взаємодії можливі лише між вузлами з однаковими номерами у загальній сукупності вузлів, то матриці $\mathbf{A}^{mk}$ мають діагональну структуру та діагональні елементи цих матриць $a_{ii}^{mk}$ можуть бути відмінними від 0 тоді і тільки тоді, коли вузол $n_i$ входить до складу $m$-го та $k$-го шару БШМ, $m \neq k$. Сумарний або агрегат-ступінь $\delta_i$ міжшарових взаємозв'язків вузла $n_i$ у БШМ загалом для багатошарових неорієнтованих мереж визначається кількістю шарів, з вузлами яких цей вузол має зв'язки.

Глобальні характеристики вузла у багатошаровій мережі, наприклад, його центральності різних типів у ТСМ також визначаються [32, 33], як вектор центральностей вузла в окремих шарах БШМ. Тобто, важливість вузла за певним показником центральності змінюється при переході від шару до шару. Такий підхід є цілком прийнятним для багатовимірних БШМС, у яких перехід потоку з шару на шар загалом є неможливим. Однак, для монопотокових частково покритих багатошарових систем, враховуючи можливість прямого або опосередкованого взаємозв'язку між будь-якими двома вузлами зв'язних структур, які породжуються міжсистемними взаємодіями такого типу, можна визначити глобальні характеристики вузлів для БШМ загалом. Так, центральність за ступенем вузла $n_i$ у частково покритій багатошаровій мережі визначається за значеннями $d_i$, $i = \overline{1, N^M}$, його агрегат-ступенів обчисленими за формулами (3). Загалом структурна модель багатошарової мережі – це не тільки опис у вигляді відповідної матриці суміжності її структури, але й сукупність основних локальних та глобальних характеристик елементів цієї структури.

# АГРЕГАТ-МЕРЕЖІ БАГАТОШАРОВИХ МЕРЕЖ

Локальною характеристикою $\varepsilon_{ij}$ ребра $(n_i, n_j)$ у БШМ, де $n_i$ та $n_j$ – вузли із загальної сукупності $V^M$, яку називатимемо його агрегат-вагою, є кількість шарів, у яких таке ребро є присутнім. Агрегат-вага $\varepsilon_{ii}$ вузла $n_i$ у частково покритій багатошаровій мережі дорівнює кількості шарів, до складу яких він входить. Вузол $n_i$, який належить кільком шарам-мережам БШМ, тобто $\varepsilon_{ii} > 1$, та через який можливий перехід потоку з одного шару на інший, називатимемо точкою переходу багатошарової мережі. Для довільної монопотокової частково покритої БШМ матриця суміжності $\mathbf{E} = \{\varepsilon_{ij}\}_{i,j=1}^{N^M}$ повністю визначає зважену мережу, яку називатимемо агрегат-мережею цієї ЧП БШМ. Елементи матриці $\mathbf{E}$ визначають інтегральні структурні характеристики вузлів та ребер багатошарової мережі (рис. 2). Для мультипотокових багатовимірних мереж значення агрегат-ваг $\varepsilon_{ij}$ зваженої агрегат-мережі визначає кількість взаємодій різних типів між вузлами таких структур. Проекція на зважену агрегат-мережу (ЗАМ) багатовимірної БШМ втрачає певну змістовність, оскільки вага кожного її ребра визначає сумарну кількість зв'язків різних типів. Для монопотокових БШМ цей недолік відсутній, оскільки вага кожного ребра відображає кількість можливих носіїв або систем-операторів, які можуть забезпечити рух відповідного типу потоку.

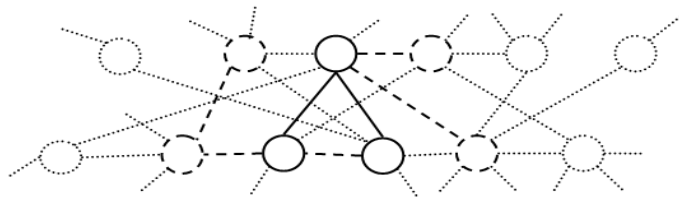

Рис. 2. Фрагмент агрегат-мережі для зображеної на рис. 1 частково покритої тришарової мережі (_____ – для $\varepsilon_{ij}=3$, _ _ _ – для $\varepsilon_{ij}=2$, ….. – для $\varepsilon_{ij}=1$)

У монопотокових ЧП БШМ зв'язки у точках переходу існують між усіма шарами, до складу яких ці точки входять (пасажир на зупинці громадського транспорту може скористатить усіма його видами, які через цю зупинку проїжджають). Очевидно, що локальні характеристики вузлів БШМ взаємопов'язані з характеристиками відповідних вузлів її агрегат-мережі. Дійсно, значення агрегат-сили взаємозв'язків вузла $n_i$ у БШМ, які обчислюються за формулою (2), також задовольняють співвідношенню

$$s_i = \sum_{\substack{j=1 \\ j \neq i}}^{N^M} \varepsilon_{ij}, \qquad (4)$$

значення агрегат-ступеню міжшарових взаємозв'язків вузла $n_i$ у БШМ задовольняють співвідношенню $\delta_i = \varepsilon_{ii} - 1$, а центральність за ступенем $d_i$ вузла $n_i$, $i = \overline{1, N^M}$, визначена за формулою (3) дорівнює кількості ненульових позадіагональних елементів $i$-го рядка матриці **E**.

Для визначення центральностей інших типів зручно використовувати поняття бінарної агрегат-мережі (БАМ) БШМ, структура якої визначається через матрицю суміжності ЗАМ шляхом заміни усіх її ненульових позадіагональних елементів значеннями 1 та обнулення значень діагональних елементів матриці **E**. Тоді центральність посередництва вузла у БШМ можна визначити, як відношення суми усіх найкоротших шляхів, які проходять через цей вузол, до суми усіх найкоротших шляхів бінарної агрегат-мережі БШМ. Центральність вузлів БШМ за власним значенням можна визначити, обчисливши власні значення матриці суміжності БАМ і т. ін. Це означає, що значення важливості вузлів у ЧП БШМ загалом буде перерозподілятися та може суттєво відрізнятися від цих значень у кожному конкретному шарі. Використання понять зваженої або бінарної агрегат-мереж ЧП БШМ не лише зменшує розмірність її вихідної моделі принаймні у $M$ разів, але й дозволяє значно ефективніше вирішувати цілу низку практично важливих проблем, а саме

1. Побудова найкоротшого шляху через БШМ, який визначається через її бінарну агрегат-мережу [34, 35] (зміна типу транспортного засобу може суттєво пришвидшити або здешевити рух пасажирів та вантажів). Коли такий шлях визначений, залишається обрати оптимальний за тим або іншим критерієм носій або систему-оператор на кожному ребрі цього шляху, агрегат-вага якого перевищує 1.

2. Пошук альтернативних шляхів руху транзитних потоків через різні мережеві шари під час ізоляції певної зони в окремому мережевому шарі (використання метро у великих містах під час заторів у години пік) [36, 37]. У цьому випадку вузли та ребра, які знаходяться в ізольованій зоні окремого шару, вилучаються зі структури ЗАМ.

3. Протидія поширенню епідемій або комп'ютерних вірусів, які завдяки міжшаровим взаємодіям можуть поширюватися значно швидше, ніж в одному шарі [38, 39]. Так, ступінь вузла у БАМ визначає кількість суміжних із ним вузлів з яких (у які) може прийти загроза інфікування; агрегат-вага $\varepsilon_{ii}$ вузла $n_i$, $i = \overline{1, N^M}$, визначає кількість шарів БШМ у яких цей вузол може сприяти поширенню інфекції.

4. Знаходження шляху з довільного вузла одного шару у будь-який вузол іншого шару, особливо якщо вони лежать поза перетином множин вузлів цих шарів. Можливість руху потоку з одного мережевого шару на інший та зворотно через точки переходу розширює доступ до вузлів, недосяжних у певному мережевому шарі та дозволяють здійснювати зв'язок між непов'язаними складовими цих шарів (наприклад, рух потоків через океани чи віддалені регіони окремих країн – північні регіони Канади, центральні регіони Австралії тощо).

## *k*-СЕРЦЕВИНИ АГРЕГАТ-МЕРЕЖ ТА *p*-СЕРЦЕВИНИ БАГАТОШАРОВИХ МЕРЕЖ

У ТСМ одним із способів визначення структурно найважливіших складових багатошарових мереж та спрощення їх моделей є поняття **k**-серцевини [40]

$$\mathbf{k} = \{k_1, k_2, ..., k_M\},$$

як поєднання $k_m$-серцевин окремих шарів БШМ. Під *k*-серцевиною складної мережі розуміють таку її підмережу, ступінь кожного вузла якої є не меншим, ніж $k$ [41]. Значення $k_m$ для різних шарів можуть відрізнятися, тому що **k**-серцевини з однаковими значеннями $k_m = k$ потенційно можуть виключати зі структури вузли і навіть окремі шари, важливі саме для організації міжсистемних взаємодій (найважливіші зі структурного погляду вузли залізничної транспортної системи зазвичай мають значно менший ступінь, ніж вузли авіаційної або водної). Загалом, **k**-серцевина визначає елементи, структурно важливі більше для шарів БШМ, а не для організації міжшарових взаємодій у ній [42].

Як було показано вище, значення багатьох характеристик елементів БШМ та її ЗАМ є рівними або взаємопов'язаними. Тоді, для визначення структурно найважливіших складових БШМ ми можемо використати поняття $k_{ag}$-серцевини її зваженої агрегат-мережі. Під $k_{ag}$-серцевиною ЗАМ ми розумітимемо її найбільшу підмережу, агрегат-сила взаємозв'язків вузлів якої є не меншими, ніж $k_{ag}$. Матриця суміжності $\mathbf{E}(k_{ag})$, яка повністю визначає структуру $k_{ag}$-серцевини зваженої агрегат-мережі ЧП БШМ, очевидним чином отримується із матриці суміжності $\mathbf{E}$ шляхом видалення рядків та стовпців, для номерів вузлів яких значення $s_i < k_{ag}$, $i = \overline{1, N^M}$.

Для вирішення проблеми визначення структурно найважливіших складових міжсистемних взаємодій у БШМ введемо поняття *p*-серцевини частково покритої багатошарової мережі

$$\tilde{G}^p = (\tilde{V}^p, \tilde{E}^p),$$

як такої її багатошарової підмережі, для кожного із вузлів $n_i$ якої виконується умова $\varepsilon_{ii} \geq p$, $2 \leq p \leq M$. Матриця суміжності *p*-серцевини ЧП БШМ очевидним чином отримується із матриці $\mathbf{A}^M$ шляхом виключення рядків та стовпців, для номерів вузлів яких значення $\varepsilon_{ii}$ є меншими, ніж *p*. Це означає, що всі зв'язки між вузлами *p*-серцевини із вихідної ЧП БШМ зберігаються. Якщо

$$p_{\max} = \max_{i,j=1,N^M}\{\varepsilon_{ij}\} = M,$$

тобто структурна $M$-серцевина багатошарової мережі $\widetilde{G}^M$ є непорожньою, то досліджувану БШМ називатимемо ядерною (рис. 3а). Очевидно, що багатошарова мережа $\widetilde{G}^M$, яку називатимемо ядром ЧП БШМ, має структуру мультиплексу та забезпечує можливість якнайшвидшого переходу з одного шару ЧП БШМ на інший. Якщо виконується умова

$$p_{\max} < M,$$

то досліджувану БШМ називатимемо без'ядерною. Прикладом ядерної ЧП БШМ є загальна транспортна система світу [43], а без'ядерної – мовна багатошарова мережа (навряд чи знайдеться хоча б одна людина, яка розмовляє всма мовами Землі [44]). Зважена агрегат-мережа $p$-серцевини багатошарової мережі (рис. 3б, $p^{ag}=p$) повністю визначається матрицею суміжності

$$\mathbf{E}_p = \{\varepsilon_{ij}^p\}_{i,j=1}^{N^M}, \quad \varepsilon_{ij}^p = \begin{cases} \varepsilon_{ij}, & \text{якщо } \varepsilon_{ii}, \varepsilon_{jj} \geq p, \\ 0, & \text{якщо } \varepsilon_{ii}, \varepsilon_{jj} < p, \end{cases} \quad i,j = \overline{1,N^M}, \quad p = 2,3,\ldots,M.$$

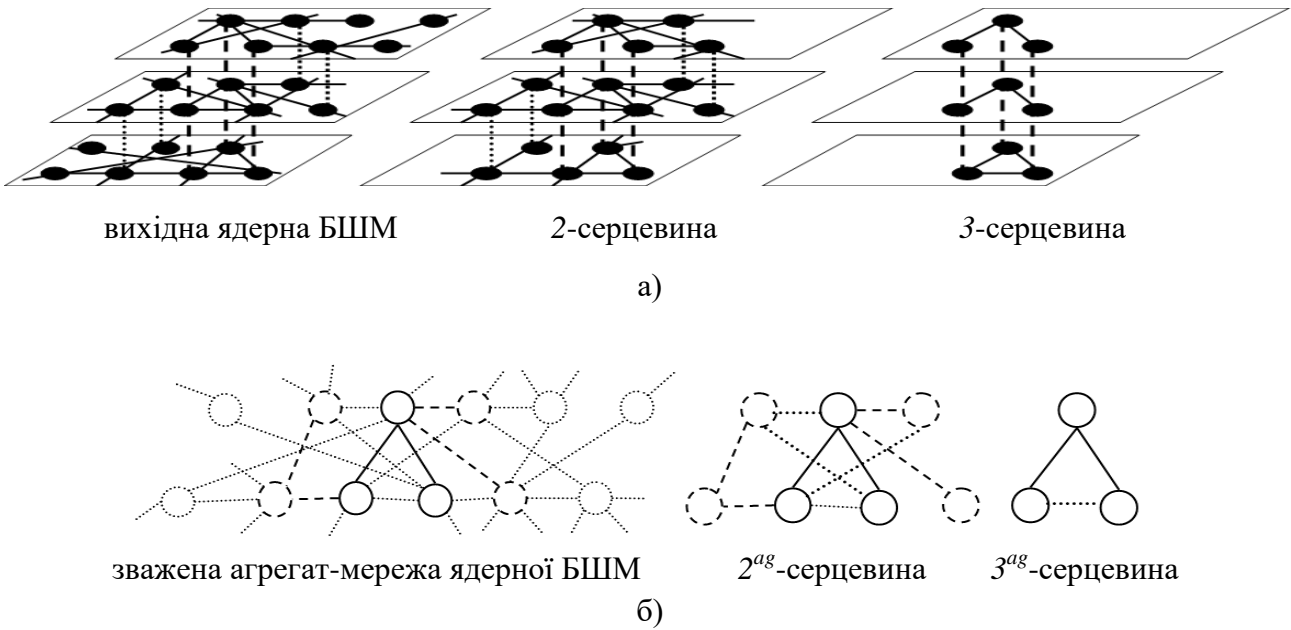

вихідна ядерна БШМ      2-серцевина      3-серцевина

а)

зважена агрегат-мережа ядерної БШМ      $2^{ag}$-серцевина      $3^{ag}$-серцевина

б)

Рис. 3. Приклади фрагментів 2- та 3-серцевин у тришаровій ядерній БШМ (а) та її зваженій агрегат-мережі (б, ⎯⎯ – елемент входить до складу трьох шарів, - - - - – елемент входить до складу двох шарів, ……. – елемент входить до складу одного шару)

Відзначимо, що $k_{ag}$-серцевина формується на підставі значень суми позадіагональних елементів рядків або стовпців матриці суміжності $\mathbf{E}$, тоді як $p^{ag}$-серцевина будується на основі діагональних елементів цієї матриці. Це означає, що безпосереднього зв'язку між цими типами серцевин не існує. Однак, їх властивості можна поєднати наступним чином. Для визначення множин вузлів з найбільшою агрегат-силою взаємозв'язків, які одночасно приймають найбільшу участь у процесі міжсистемних взаємодій, введемо поняття $p^{ag}(k_{ag})$-серцевини як такої підмережі вихідної ЗАМ, вузли якої, по-перше, мають агрегат-силу взаємозв'язків вузлів $s_i \geq k_{ag}$ і, по-друге, входять до складу не менше ніж $p$ шарів БШМ. Тут загалом $k_{ag} \neq p^{ag} = p$. Матриця суміжності $\mathbf{E}(p^{ag}(k_{ag}))$ відповідної підмережі вихідної БШМ очевидним чином отримується з матриці суміжності $\mathbf{E}(k_{ag})$ шляхом виключення рядків та стовпців, для яких діагональні елементи є меншими ніж $p$. Зворотно, для визначення множин вузлів, які приймають найбільшу участь у процесі міжсистемних взаємодій і одночасно мають найбільші значення агрегат-сили взаємозв'язків вузлів, введемо поняття $k_{ag}(p^{ag})$-серцевини, як такої підмережі вихідної ЗАМ, вузли якої, по-перше, входять до складу щонайменше $p$ шарів БШМ і, по-друге, мають агрегат-силу взаємозв'язків вузлів $s_i \geq k_{ag}^{\varepsilon}$. Матриця суміжності $\mathbf{E}(k_{ag}^{\varepsilon}(p_{ag}))$ відповідної підмережі вихідної БШМ очевидним чином отримується із матриці суміжності $p$-серцевини $\mathbf{E}_p$, $p = 2,3,...,M$, шляхом виключення рядків та стовпців, для яких агрегат-сили взаємозв'язків вузлів $s_i < k_{ag}$. Очевидно, що структури $p^{ag}(k_{ag})$- та $k_{ag}(p^{ag})$-серцевин, як і вигляд їх матриць суміжності $\mathbf{E}(p^{ag}(k_{ag}))$ та $\mathbf{E}(k_{ag}(p^{ag}))$ можуть відрізнятися, оскільки методи та цілі формування цих серцевин загалом є різними.

## СЦЕНАРІЇ ГРУПОВИХ ЦІЛЕСПРЯМОВАНИХ АТАК НА БАГАТОШАРОВІ МЕРЕЖІ НА ПІДСТАВІ СТРУКТУРНОЇ МОДЕЛІ

Уразливість окремих мережевих шарів БШМ до цілеспрямованих атак може бути визначена за допомогою сценаріїв, які базуються на послідовному вилученні вузлів мережі з найбільшими значеннями центральностей за ступенем [45], посередництвом [46], близькості [47], власних значень [48] тощо або їх поєднання [49]. Уразливість міжсистемних взаємодій до цілеспрямованих атак у монопотокових ЧП БШМ в основному визначається уразливістю точок переходу багатошарової мережі, тобто її $p$-серцевин з різними значеннями $p$, починаючи із найбільшого $p \leq M$. Вище вже зазначалося, що використання поняття агрегат-мережі БШМ дозволяє суттєво спрощувати розв'язання низки практично важливих задач системного аналізу. Не менш ефективним є його застосування для побудови дієвих сценаріїв послідовних групових атак на структуру БШМС, які базуються на інтегральних показниках важливості вузлів багатошарової мережі, а саме локальних та глобальних характеристиках елементів зважених та бінарних агрегат-мереж. Відзначимо, що видалення певного вузла зі структури системи зазвичай призводить до перерозподілу маршрутів руху потоків мережею та встановленню нових зв'язків між вузлами, що залишились. Цей процес відображає реакцію системи до змін умов функціонування та означає зміну значень центральностей різних типів вузлів БШМ [50, 51].

Перед побудовою сценаріїв цілеспрямованих атак необхідно визначити критерії успішності цих атак, тобто очікуваний рівень ураження структури системи. Такі критерії включають поділ агрегат-мережі на незв'язні складові при якому БШМ перестає існувати як єдина надсистемна формація, обмеження або припинення міжсистемних взаємодій між мережевими шарами (зменшення значення $p_{\max}$), суттєве зменшення розміру найбільшої зв'язної компоненти або середньої довжини геодезичного шляху [49] тощо. Побудуємо перший сценарій цілеспрямованої послідовної групової атаки, використовуючи поняття агрегат-ваг та агрегат-сили взаємозв'язків вузлів зваженої агрегат-мережі БШМ:

1) створимо перелік вузлів зваженої агрегат-мережі БШМ у порядку зменшення їх агрегат-ваг (значень $\varepsilon_{ii}$ матриці **E**);

2) у кожній групі вузлів з однаковими значеннями $\varepsilon_{ii}$ впорядковуємо вузли за зменшенням значень їх агрегат-сили взаємозв'язків $s_i$, обчисленої за формулою (2);

3) видаляємо перший вузол зі створеного списку; якщо обраний критерій успішності атаки досягнуто, то виконання алгоритму завершено, інакше переходимо до наступного пункту;

4) видалення певного вузла звично призводить до встановлення нових зв'язків між вузлами, які залишились у БШМ, тобто структура вузлів у різних шарах БШМ та її зваженої агрегат-мережі та, відповідно, значень $s_i$ вузлів може змінитися; тоді, якщо перелік вузлів з поточним значенням $\varepsilon_{ii}$ не вичерпаний, то переходимо до пункту 2 цього сценарію, інакше – до наступного кроку;

5) після опрацювання усіх вузлів групи з поточним значення $\varepsilon_{ii}>1$ переходимо до кроку 2 зі значенням $\varepsilon_{ii}-1$; якщо $\varepsilon_{ii}=1$, то виконання алгоритму завершено.

Другий сценарій послідовної групової цілеспрямованої атаки будуємо на підставі використання агрегат-ваг та центральності посередництва вузлів бінарної агрегат-мережі БШМ:

1) створимо перелік вузлів зваженої агрегат-мережі БШМ у порядку зменшення їх агрегат-ваг (значень $\varepsilon_{ii}$ матриці **E**);

2) у кожній групі вузлів з однаковими значеннями $\varepsilon_{ii}$ впорядковуємо вузли на підставі зменшення значень їх центральності посередництва у бінарній агрегат-мережі БШМ;

3) видалимо перший вузол з початку створеного списку; якщо критерій успішності атаки виконано, то виконання алгоритму завершено, інакше переходимо до наступного кроку;

4) видалення певного вузла звично призводить до встановлення нових зв'язків між вузлами, які залишились у БШМ, тобто структура вузлів у різних шарах БШМ та її бінарній агрегат-мережі та, відповідно, значень центральності посередництва вузлів може змінитися; тоді, якщо перелік вузлів з поточним значенням $\varepsilon_{ii}$ не вичерпаний, то переходимо до пункту 2 цього сценарію, інакше – до наступного кроку;

5) після опрацювання усіх вузлів групи з поточним значення $\varepsilon_{ii}>1$ переходимо до кроку 2 зі значенням $\varepsilon_{ii}-1$; якщо $\varepsilon_{ii}=1$, то виконання алгоритму завершено.

Зазвичай сценарії, які перераховують характеристики елементів після видалення чергового вузла та базуються на центральності посередництва, є більш ефективними для досягнення цілі атаки, ніж сценарії без перерахунку та базовані на центральності за ступенем або агрегат-сили взаємозв'язків вузлів БШМ [46]. Аналогічно можна побудувати сценарії послідовних групових цілеспрямованих атак на ЗАМ або БАМ багатошарової мережі, базуючись на інших показниках центральності вузлів агрегат-мережі БШМ.

Під одночасною, на відміну від послідовної, ми розуміємо таку атаку на БШМ, під час якої відвовідна багатошарова система не в змозі перерозподілити функції уражених елементів між тими, які залишились неураженими (встановити нові зв'язки між ними). Захиститися від успішної одночасної атаки на групу найважливіших елементів БШМ і, головне, подолати наслідки такої атаки набагато складніше [52, 53]. Для ураження міжсистемних взаємодій такою групою повинна бути певна сукупність найважливіших точок переходу. Поняття $p^{ag}$-, $p^{ag}(k_{ag})$- та $k_{ag}(p^{ag})$-серцевин дають змогу визначити найважливіші зі структурного погляду групи вузлів БШМ, одночасна цілеспрямована атака на які призведе до найбільшого ураження або навіть знищення структури міжсистемних взаємодій. Оскільки принципи побудови сценаріїв одночасних групових атак на перелічені вище серцевини БШМ є подібними, то опишемо сценарій такого ураження, базуючись на понятті $k_{ag}(p^{ag})$-серцевини, який зводиться до послідовного виконання наступних кроків:

1) приймаємо рівним $p = p^{\max} \leq M$ та $k_{ag} = \max\limits_{i=1,N^M} s_i$ ;

2) видаляємо зі структури БШМ вузли, які входять до складу $k_{ag}(p^{ag})$-серцевини; якщо обраний критерій успішності атаки виконано, то виконання алгоритму завершено, інакше переходимо до наступного пункту;

3) зменшуємо значення $k_{ag}$ на 1; якщо перелік вузлів $k_{ag}(p^{ag})$-серцевини не вичерпаний, то переходимо до пункту 2, інакше виконуємо наступний крок;

4) якщо значення $p>1$, то зменшуємо його на 1 та переходимо до пункту 2 зі значенням $k_{ag}$, рівним максимальній агрегат-силі взаємозв'язків вузлів, які залишились в агрегат-мережі; інакше завершуємо виконання алгоритму.

**ЗАГАЛЬНОСИСТЕМНІ ЦІЛЕСПРЯМОВАНІ АТАКИ НА БАГАТОШАРОВІ МЕРЕЖІ**

Під загальносистемною ми розуміємо таку атаку на ЧП БШМ, яка повністю вражає структуру одного або кількох її шарів-систем. Прикладами таких атак можна вважати блокування під час російсько-української війни пасажирського авіаційного та вантажного водного шарів загальної транспортної системи України. При цьому, якщо заборону пасажирських авіаперевезень вдалося принаймні частково компенсувати зарахунок інших видів транспорту, зокрема залізничного та автомобільного, то блокування морських портів призвело до припинення значної частки вантажних перевезень аграрної та металургійної продукції українських виробників. Перенаправлення їх залізничним транспортом вдалося здійснити у дуже незначних обсягах із-за обмеженої пропускної здатності залізничних шляхів. Загалом, блокування морського шару є прикладом того, як ураження одного із шарів БШМ призводить до дестабілізації як внутрішньо, так і міжшарових взаємодій у відповідній багатошаровій системі [54]. Блокування унаслідок санкцій, які були застосовані до країни-агресора, її банківської сфери та низки важливих галузей промисловості можна вважати прикладами ефективної загальносистемної атаки на фінансово-економічну систему рф [55]. Однак, послідовне застосування окремих пакетів санкцій зробило цю атаку менш дієвою, ніж якби вони були впроваджені одночасно.

На підставі структурної моделі багатошарової мережі (1) ми можемо визначити характеристики шарів БШМ, які дають змогу встановити першочергові зі структурного погляду цілі атаки, а саме [56]

1) питому вагу $\theta_m$ сукупності вузлів $m$-го шару у загальній сукупності вузлів частково покритої БШМ, яка визначається співвідношенням

$$\theta_m = N_m / N^M,$$

де $N_m$ – кількість вузлів $m$-го шару;

2) питому вагу $\vartheta_m$ сукупності зв'язків $m$-го шару у загальній сукупності зв'язків частково покритої БШМ, яка визначається співвідношенням

$$\vartheta_m = L_m / L^M,$$

де $L_m$ – кількість бінарних зв'язків $m$-го шару, $L^M$ – кількість загальної сукупності ребер бінарної агрегат-мережі БШМ;

3) питома вага $\psi_m$ точок переходу $m$-го шару у сукупності всіх точок переходу частково покритої БШМ, яка визначає доступність міжшарових взаємодій для даного шару та обчислюється як відношення кількості діагональних елементів блоку $\mathbf{A}^{mm}$, $m = \overline{1, M}$, матриці $\mathbf{A}^M$, значення яких є більшими ніж 1 до кількості діагональних елементів матриці $\mathbf{A}^M$, значення яких є також більшими 1.

Сценарії послідовних атак на шари БШМ будуються за аналогічними описаним вище принципами, тобто спочатку здійснюється впорядкування послідовності шарів у порядку зменшення значень обраної із визначених вище характеристик важливості шару у БШМ, а потім здійснюється атака на перший із шарів цього списку. Далі, враховуючи, що із вилученням певного шару структура міжсистемних взаємодій може змінитися, перелік важливості шарів корегується та здійснюється атака на перший із шарів новоствореного списку. Цей алгоритм продовжується поки не буде виконаний обраний критерій успішності атаки. Сценарії одночасних цілеспрямованих атак на кілька або всі шари БШМ базуються на визначенні сукупності шарів із найвищими показниками важливості та подальшому ураженні шарів із цієї сукупності. Прикладами таких одночасних атак є вже згадані вище пакети фінансово-економічних санкцій проти рф, як країни-агресора.

Для загальної транспортної системи України питома вага сукупності вузлів окремих шарів приймає наступні значення: $\theta_{авто} \approx 1{,}0$, $\theta_{залізн} \approx 0{,}04$, $\theta_{авіа} \approx 0{,}000625$ та $\theta_{водний} \approx 0{,}00025$, а питома вага точок переходу окремих шарів у сукупності всіх точок переходу – значення $\psi_{авто} \approx 1{,}0$, $\psi_{залізн} \approx 1{,}0$, $\psi_{авіа} \approx 0{,}0156$ та $\psi_{водний} \approx 0{,}007$ [56]. Однак, як показує досвід російсько-української війни, уразити найважливіші складові автомобільної та залізничної транспортних систем-шарів достатньо складно. Це пояснюється не тільки кількістю елементів цих МС, але й швидкістю їх відновлення (для залізниці – до 72 год.). Заборона заради безпеки пасажирів авіаперельотів хоча й створює певні незручності, але достатньо просто заміщається залізничним та автомобільним транспортом. Однак, блокування водного шару загальної транспортної системи України створило суттєві труднощі для експорту аграрної та металургійної продукції українських виробників і значно скоротило обсяги вантажних залізничних та автоперевезень. Очевидно, що структурні показники, принаймні у цьому випадку, недостатньо адекватно відображають важливість шарів транспортної багатошарової мережевої системи. Як було показано у [1, 4], на підставі потокових моделей МС можна фор-мувати значно адекватніші з функціонального погляду показники такої важливості та будувати суттєво дієвіші сценарії як групових, так і загальносистемних атак на БШМС.

**ВИСНОВКИ**

У статті на підставі структурної моделі багатошарової мережевої системи побудована модель її агрегат-мережі, яка дає змогу суттєво зменшувати розмірність низки практично важливих задач дослідження структур міжсистемних взаємодій та визначати інтегральні показники важливості вузлів БШМС. Введено поняття $k_{ag}$-серцевини агрегат-мережі та $p$-серцевини структури міжсистемних взаємодій, як найважливіших зі структурного погляду складових багатошарових мереж, ураження яких може призвести до дестабілізації процесу функціонування як окремих шарів-систем, так і процесу міжсистемних взаємодій загалом. Виділення таких складових БШМ сприяє подоланню проблеми складності системних досліджень, зокрема, зменшенню розмірності їх моделей, визначенню інтегральних показників важливості елементів структури, кращому розумінню процесів розвитку міжсистемних взаємодій тощо. Побудовані сценарії послідовних та одночасних групових, а також загальносистемних атак на структуру БШМС, які дають змогу визначати ті її складові, які потребують першочергового захисту. Однак, ураження процесу функціонування системи можливе і за неураженої структури. До того ж на рівні структурного підходу достатньо важко проаналізувати наслідки уражень, заподіяних цілеспрямованою атакою, та розробити стратегії відновлення процесу функціонування уражених елементів, підсистем або системи загалом. Тому наступним кроком наших досліджень є розроблення на підставі потокових моделей складних мережевих систем та міжсистемних взаємодій сценаріїв цілеспрямованих групових та загально системних атак на найважливіші з функціонального погляду складові БШМС.

**ПЕРЕЛІК ПОСИЛАНЬ**